\documentclass[twocolumn]{aastex62}
\usepackage{amsmath}
\usepackage{amsfonts}

\submitjournal{ApJ}

\usepackage[]{color}
\usepackage{bm}        
\usepackage{tabularx}        
\newcolumntype{Y}{>{\centering\arraybackslash}X}

\newcommand{\pitch}{\tilde{\mu}}

\newcommand{\vect}[1]{{\bm #1}}
 
\begin{document}

\title{Safety first: Stability and dissipation of line-tied force-free flux tubes in magnetized coronae}
\shorttitle{Stability of line-tied force-free flux tubes}
\shortauthors{Rugg et al.}

\correspondingauthor{N. Rugg}
\email{nataliejrugg@gmail.com}

\author[0000-0002-3778-043X]{N. Rugg}
\affiliation{Independent researcher}
\altaffiliation[Science Undergraduate Laboratory Internship (SULI) at the\\Princeton Plasma Physics Laboratory (PPPL)]{}

\author[0000-0002-5349-7116]{J. F. Mahlmann}
\affiliation{Department of Astrophysical Sciences, Peyton Hall, Princeton University, Princeton, NJ 08544, USA}
\affiliation{Department of Astronomy \& Astrophysics, Pupin Hall, Columbia University, New York, NY 10027, USA}

\author[0000-0001-9179-9054]{A. Spitkovsky}
\affiliation{Department of Astrophysical Sciences, Peyton Hall, Princeton University, Princeton, NJ 08544, USA}


\begin{abstract}
Magnetized plasma columns and extended magnetic structures with both footpoints anchored to a surface layer are an important building block of astrophysical dissipation models. Current loops shining in X-rays during the growth of plasma instabilities are observed in the corona of the Sun and are expected to exist in highly magnetized neutron star magnetospheres and accretion disk coronae. For varying twist and system sizes, we investigate the stability of line-tied force-free flux tubes and the dissipation of twist energy during instabilities using linear analysis and time-dependent force-free electrodynamics simulations. Kink modes ($m=1$) and efficient magnetic energy dissipation develop for plasma safety factors $q\lesssim 1$, where $q$ is the inverse of the number of magnetic field line windings per column length. Higher-order fluting modes ($m>1$) can distort equilibrium flux tubes for $q>1$ but induce significantly less dissipation. In our analysis, the characteristic pitch $\pitch_0$ of flux tube field lines determines the growth rate ($\propto \pitch_0^3$) and minimum wavelength of the kink instability ($\propto \pitch_0^{-1}$). We use these scalings to determine a minimum flux tube length for the growth of the kink instability for any given $\pitch_0$. By drawing analogies to idealized magnetar magnetospheres with varying regimes of boundary shearing rates, we discuss the expected impact of the pitch-dependent growth rates for magnetospheric dissipation in magnetar conditions.
\end{abstract}

\received{December 7, 2023}
\revised{February 20, 2024}
\accepted{February 20, 2024}

\section{Introduction}

Magnetospheric dissipation likely drives at least some of the abundant X-ray activity observed from compact object magnetospheres with active coronae, such as magnetars \citep[e.g.,][]{Gogus1999,Gogus2000,Gogus2001,Rea2009,Rea2010,Kaspi_2017ARAaA..55..261,Esposito2020} and magnetized black hole accretion disks \citep[e.g.,][]{Haardt1994,Matteo1999,Chartas2009,Uttley2014,Wilkins2015}. The stability and dynamics of flux bundles in (highly) magnetized environments are well-studied and observed, including for applications to astrophysical jets \citep[e.g.,][]{Lyubarskii1999,Giannios2006,Lapenta2006,Narayan2009,Alves2018,Bromberg2019,Davelaar2020} and the solar corona \citep[e.g.,][]{Raadu1972,Hood1979,Hood1981,Linton1998,Lapenta2006,Kumar2014,FloridoLlinas2020,Xu2020,Quinn2022}. Yet, insights from the considered systems, often periodic or highly constrained by the specific astrophysical scenario, cannot answer the most fundamental questions for highly magnetized flux tubes with field line footpoints frozen (line-tied) to a stellar or disk surface boundary at both ends: \emph{If, when, and how does a flux tube become unstable and what is the amount of dissipated magnetic energy during its instability?}

This work combines insights from different fields of plasma astrophysics. First, we exploit the vast literature on magnetic flux rope dynamics in the solar corona \citep[e.g.,][]{Galsgaard1997,Amari2003,Gerrard2004,Torok2005,Torok2010,Gordovskyy2011,Gordovskyy2014,Pinto2016,Ripperda2017,Ripperda2017a,Sauppe2018}, where the injection of twist and helicity from surface motion produces a variety of dissipative events. Second, we use well-established constraints from laboratory plasma physics \citep[e.g.,][]{hazeltine2003plasma,Bergerson2006,longaretti2008}. The so-called safety factor denotes the inverse of the ratio of field line windings per column length and indicates the susceptibility of a flux tube to plasma instabilities \citep[e.g.,][]{Goedbloed1972,goedbloed2019}. Third, we repurpose numerical methods from the study of relativistic jets. Growth rates of perturbations to rotating equilibrium flux tubes of infinite length were derived numerically by \citet{Sobacchi2017}, and we closely follow their implementation and analysis. Finally, we use the results from three-dimensional numerical models of a global magnetar magnetosphere to evaluate the astrophysical implications of our findings. \citet{Carrasco:2019aas} and then with broader parameter ranges. \citet{Mahlmann2023} find eruption scenarios that range in onset time and dissipation for flux tubes twisted at one end by surface motions. However, their work does not explore a reliable instability criterion on the eruption of three-dimensional twisted flux tubes. We analyze the plasma safety factor as an instability criterion for coronal loops around compact astrophysical objects by drawing analogies to well-established theories. For example, \citet{Raadu1972} and \citet{Hood1979,Hood1981} studied the stabilizing effect of line-tying for magnetic columns with critical safety factors to explain the energy release of coronal mass ejections (CMEs). Solar flux tubes pinch regions of weak magnetic fields with shallow gradients at their edges. In contrast, magnetar current loops can have sharp edges on a strong background field, with negligible contributions by plasma pressure gradients. Their magnetic field lines are tied to a perfectly conducting stellar surface, while the ends of a solar flux tube extend through the photosphere and chromosphere.

Force-free electrodynamics (FFE), the vanishing-inertia limit of ideal magnetohydrodynamics (MHD), is a good approximation for modeling the global dynamics of highly magnetized magnetospheric plasma \citep[see, e.g.,][]{gruzinov1999,Blandford2002,Komissarov2004,Spitkovsky2006,Parfrey2012,Carrasco2017,Most2020,yuan2020,ripperda2021}. FFE methods gain efficiency by disregarding the exact physics of nonideal dissipation, namely the screening of electric fields $E_\parallel$ along the magnetic field or in electrically dominated regions ($E>B$). However, especially with high-order numerical techniques, one can capture with good accuracy the evolution of magnetic pressure and tension as well as (non)linear interactions of plasma modes. In this work, we use FFE to model the growth of instabilities in perturbed flux tube equilibria with static line-tied boundaries. We probe an instability criterion for the onset of the kink mode and give limits on the dissipated magnetic energy for different evolution scenarios. 

This paper is organized as follows. Section~\ref{sec:equilibria} semianalytically studies the linear growth of perturbations to force-free equilibrium flux tubes. Section~\ref{eq:sims} validates the expected instability growth (Section~\ref{sec:dynamics}) and analyzes the dissipation of twist energy for various flux tube parameters (Section~\ref{sec:twistenergy}). Our discussion in Section~\ref{sec:discussion} provides scalings and limits of the instability growth (Section~\ref{sec:discussionscales}), applies our findings to magnetized astrophysical coronae (Sections~\ref{sec:discussionmagnetar} and~\ref{sec:discussionmixed}), and discusses limitations (Section~\ref{sec:limitations}). We state conclusions in Section~\ref{sec:conclusions} and give additional details on the instability evolution in Appendix~\ref{app:relaxedstates}.

\section{Force-free flux tube equilibria}
\label{sec:equilibria}

\begin{figure}
    \centering
  \includegraphics[width=\linewidth]{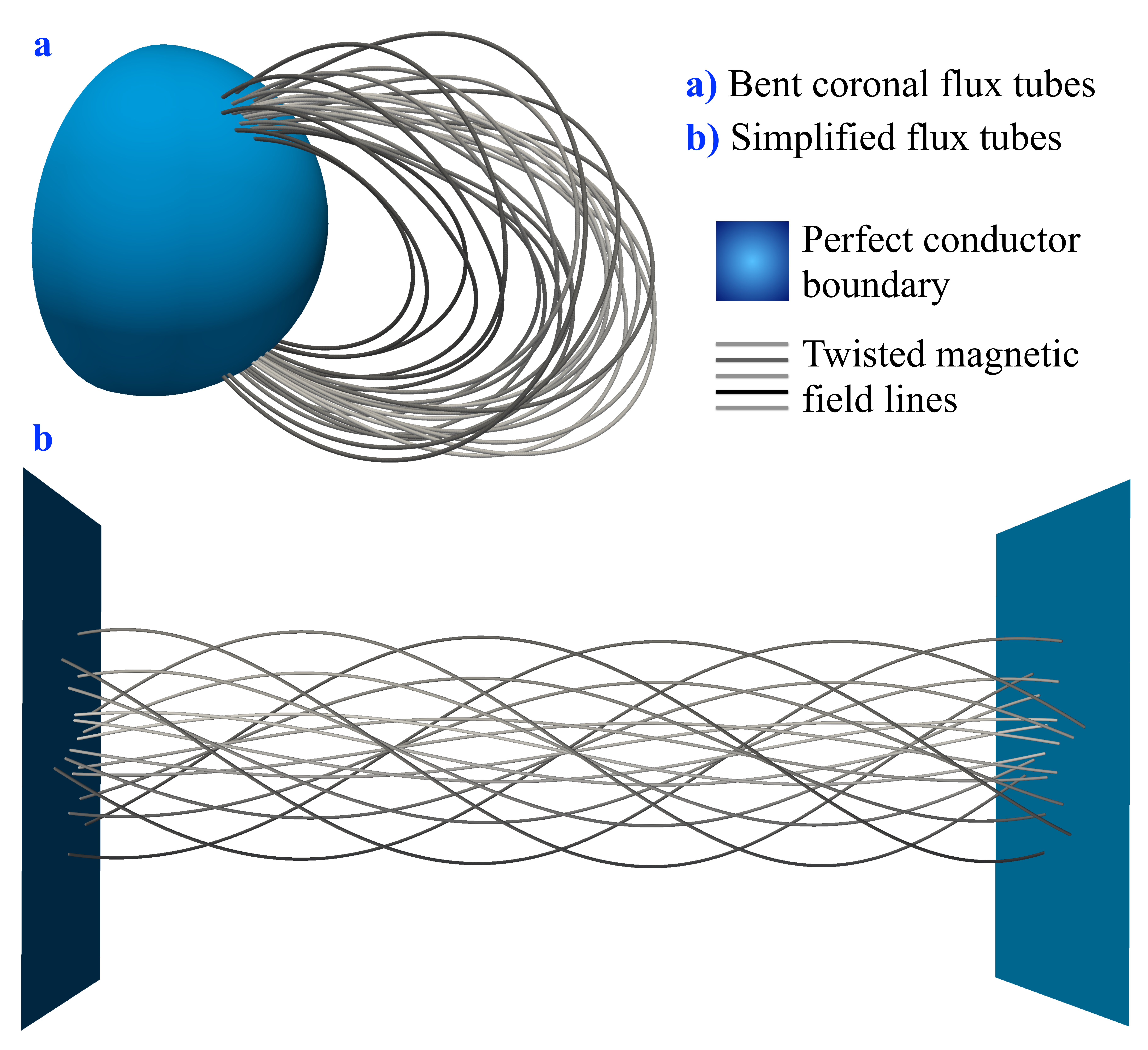}
  \caption{Schematic illustration of the configurations studied in this paper (panel b) and their astrophysical context (panel a). Coronal flux ropes, as observed on the Sun and expected around magnetars and magnetized accretion disks, are usually bent (panel a) with magnetic field lines frozen to a surface layer. We mimic such surface layers by perfectly conducting surfaces as boundary conditions in our simulations. Studying the stability of bent flux tubes on arbitrary background magnetic fields is not straightforward. We, therefore, analyze simplified flux tubes as twisted magnetic field lines embedded in a uniform background magnetic field (panel b).}
  \label{fig:schematics}
\end{figure}

Formed by twisted magnetic field lines arched with footpoints frozen into a surface layer, coronal flux ropes are expected in several astrophysical systems, such as stars and magnetized accretion disks (see introductory references). Surface motions can drag along the line-tied magnetic field lines and disrupt the magnetized flux tube equilibria. To study the stability of highly magnetized twisted magnetic fields, we analyze simplified flux tube geometries embedded in uniform background magnetic fields (see Figure~\ref{fig:schematics}). We evaluate Maxwell's equations,
\begin{align}
    c\nabla\times\vect{E}&=-\partial_t\vect{B}\label{eq:MAXI}\,,\\
    \nabla\cdot\vect{E}&=4\pi\rho\,,\\
    c\nabla\times\vect{B}&=4\pi\vect{j}+\partial_t\vect{E}\,,\label{eq:MaxIII} \\
    \nabla\cdot\vect{B}&=0\,,\label{eq:MAXIV}
\end{align}
for ideal electric fields with frozen-in magnetic flux
\begin{align}
    \vect{E}=-\frac{\vect{v}}{c}\times\vect{B}\label{eq:idealeq},
\end{align}
and for a vanishing Lorentz force
\begin{align}
\rho\vect{E}+\frac{\vect{j}}{c}\times\vect{B}=\vect{0}\label{eq:ffbalance}.
\end{align}
For stationary electromagnetic fields, Equation~(\ref{eq:ffbalance}) becomes the equilibrium condition:
\begin{align}
    \left(\nabla\cdot\vect{E}\right)\vect{E}+\left(\nabla\times\vect{B}\right)\times\vect{B}=\vect{0}
    \label{eq:ffbalanceEB}.
\end{align}
For cylindrical coordinates $(r,\phi,z)$, magnetic field lines can move around the symmetry axis with a velocity $v_{F\phi}=r \Omega_{\rm F}$, where $\Omega_{\rm F}$ is the field line angular velocity. In stationary and axisymmetric configurations, the only nonvanishing component of the electric field is $E_r=-r\Omega_{\rm F}B_z/c$ (Equation~\ref{eq:idealeq}). The angular velocity and corresponding electric field are conserved along flux surfaces. In this geometry, the radial component of Equation~(\ref{eq:ffbalanceEB}) yields the generalized force-free Grad-Shafranov equation of a flux tube\footnote{The Grad-Shafranov equation balances magnetic tension and pressure with the plasma pressure gradient in toroidal MHD equilibria \citep[][and references therein]{goedbloed2019}. This work studies the \emph{force-free} limit of such equilibria, absent of inertial properties like plasma density, velocity, and pressure. It also \emph{generalizes} them to cylindrical coordinates.}:
\begin{align}
    \partial_r\left[\frac{r^2\Omega_{\rm F}B_z}{c}\right]\frac{\Omega_{\rm F}B_z}{c}=B_z\partial_r B_z+\frac{B_\phi}{r}\partial_r \left[r B_\phi\right]\label{eq:GSBasic}
\end{align}
We express the magnetic field $B_\phi$ in the orthonormal basis. This work studies line-tied force-free equilibria with footpoints anchored to perfectly conducting boundaries and a vanishing field line angular velocity $\Omega_F$. In this limit, Equation~(\ref{eq:GSBasic}) becomes
\begin{align}
\begin{split}
rB_z\partial_r B_z=-B_\phi\partial_r B_\phi
\;\;\Rightarrow\;\; r\frac{B_z}{B_\phi}=-\frac{\partial_r \left[rB_{\phi}\right]}{\partial_r B_{z}}.
\end{split}\label{eq:staticbalance}
\end{align}
We can define the \emph{inverse} pitch parameter $\pitch=B_\phi/B_z$ and write Equation~(\ref{eq:staticbalance}) as
\begin{align}
    1=-\pitch^2-\left(\frac{\pitch^2}{r}+\pitch\;\partial_r \pitch\right)\frac{B_z}{\partial_r B_{z}}.
    \label{eq:staticpitched}
\end{align}
This equation determines the equilibrium magnetic fields of a static flux tube without field line rotation for radial profiles of the pitch $\pitch(r)$. It is well studied throughout the literature in Newtonian ideal MHD \citep[e.g.,][]{goedbloed2019,goldston2020} and has exact solutions in some cases like uniform twists \citep{Gold1960} or certain oscillating magnetic fields \citep{lundquist1950}. In this work, we study field line columns similar to those twisted by footpoints moving on a stellar surface or an accretion disk. We choose simple pitch profiles to capture two main properties: the rapid decay of twist at flux tube boundaries, and a twist profile compatible with certain surface motions. In the following, we evaluate the stability of flux tubes with
\begin{align}
    \pitch(r)=\pitch_0 r^\nu f(r)\label{eq:pitch},
\end{align}
where $\nu\geq 1$. For a characteristic length scale $r_0$, we choose $f(r) = 0.5\times\tanh[(r/r_0-a)/b]$ with $a\approx 1.11$ and $b=0.1$. This choice induces a pitch profile with a maximum at $r/r_0=1$, vanishing for $r\gg r_0$. For the following instability analysis, we use $\nu\in\left\{1,2\right\}$ to probe different pitch profiles in the flux tube. By integrating Equation (\ref{eq:staticpitched}) we generate equilibrium magnetic fields for boundary conditions $B_z(r\gg r_0)=B_{\rm bg}$, where $\mathbf{B}_{\rm bg}=B_{\rm bg}\mathbf{\hat{z}}$. We then use solutions to Equation~(\ref{eq:staticpitched}) as \emph{background fields} in an instability analysis of helical force-free MHD equilibria \citep{Solovev1967,Lyubarskii1999,Sobacchi2017}.

\subsection{Instability analysis}
\label{sec:instability}

We use the numerical method introduced by \citet{Sobacchi2017} to find growth rates for linear instabilities of flux tubes with inverse pitch profiles given by Equation~(\ref{eq:pitch}). We numerically derive growth rates of linear perturbations $\xi$ of the form
\begin{align}
    \xi\propto \xi(r)\times e^{\text{i}\left(\omega t +m\phi-k z\right)}.
\end{align}
Here, $\xi(r)$ is the complex-valued perturbation along the flux tube radius with frequency $\omega$ and wavenumbers $(m,k)$ along the $\phi$ and $z$-directions, respectively. The vertical wavelength associated with such a perturbation is given by $\lambda=2\pi/k$ and its growth rate by the imaginary contribution $\text{Im}(\omega)$. In practice, we follow \citet{Sobacchi2017} and discretize a complex-valued balance equation for perturbations $\xi$ to helical force-free equilibria \citep[][Equation~20]{Lyubarskii1999} on a one-dimensional mesh along the radial direction of the flux tube\footnote{The technique of linearizing balance laws like Equation~(\ref{eq:GSBasic}) is common in the Newtonian MHD literature \citep[e.g.,][]{hain1958,Frieman1960,goedbloed1971}. 
For certain background fields, growth rates of the kink mode (like Figure~\ref{fig:M1GROWTH}) were derived early on \citep[e.g.,][]{Goedbloed1972}. The strategy adopted by \citet{Lyubarskii1999} and \citet{Sobacchi2017} and employed in this work follows the analysis of nonrelativistic MHD equations for helical stationary flows by \citet{Solovev1967}. We refer the reader to these works for a vast background of the earlier theoretical development.}. We impose radial boundary conditions $\xi'(0)=0$ with an arbitrary normalization $\xi(0)=1$, as well as $\xi(r\gg r_0)=0$ \citep[see Section~2.2 in][]{Sobacchi2017}. We neglect line-tying boundaries of the perturbation $\xi$ along the $z$-direction (see limitation in Section~\ref{sec:limitations}). For \emph{freely chosen} wavenumbers $(m,k)$, an initially \emph{estimated} complex frequency $\omega$ is then driven to a solution of the balance equation by minimizing the residual error of the shooting method \citep[e.g.,][]{Vetterling1992}. 

\subsubsection{The $m=1$ (kink) mode}

\begin{figure}
    \centering
  \includegraphics[width=\linewidth]{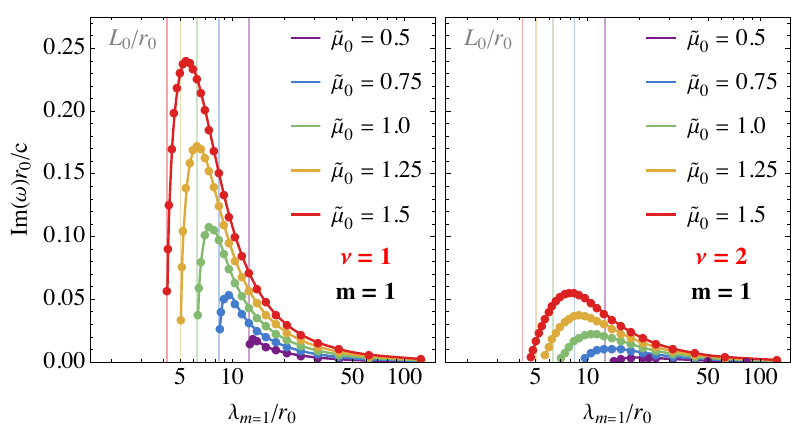}
  \caption{Growth rates of the $m=1$ mode for varying pitch and profiles of the twisted flux tube (Equation~\ref{eq:pitch}) as a function of the unstable mode wavelength $\lambda$. Circles indicate the numerically derived rates for $\nu\in\left\{1,2\right\}$ and $\pitch_0\in\left\{0.5,0.75,1.0,1.25,1.5\right\}$. We indicate the critical length of safety factor $q=1$ by vertical lines for different pitch parameters $\pitch_0$.}
  \label{fig:M1GROWTH}
\end{figure}

\begin{figure}
\centering
  \includegraphics[width=\linewidth]{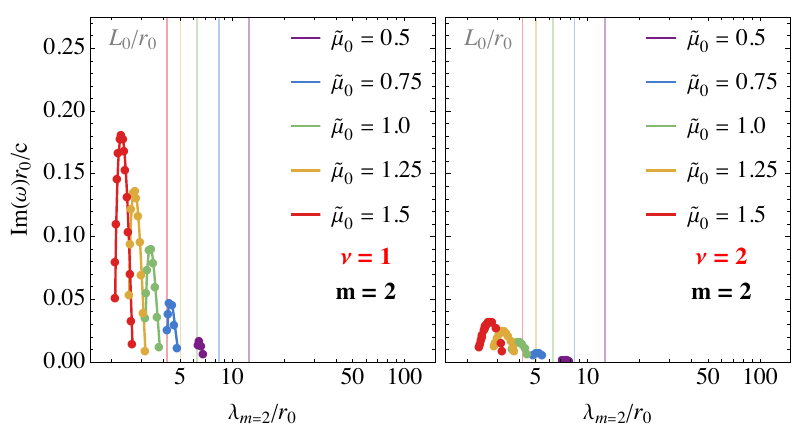}
  \caption{Growth rates of the $m=2$ (fluting) mode, as Figure~\ref{fig:M1GROWTH}.}
  \label{fig:M2GROWTH}
\end{figure}

We first quantify the dynamics of the fastest-growing nonaxisymmetric instability of the flux tubes, that is, the kink mode. A commonly employed measure of the susceptibility of magnetic columns to kink instability is the so-called safety factor $q$. This parameter represents the inverse of the number of magnetic field line windings distributed along the tube length $L$:
\begin{align}
    q\equiv\frac{2\pi r_0 }{L} \frac{B_z}{B_\phi} =\frac{2\pi r_0 }{L\pitch}\,.
    \label{eq:safetyfactor}
\end{align}
In the setup given by Equation~(\ref{eq:pitch}), the safety factor is minimal at $r/r_0=1$. An instability is expected for $q\lesssim 1$, and for each radial pitch profile we define the critical length corresponding to this threshold as $L_0=2\pi r_0/\pitch_0$.

In Figure~\ref{fig:M1GROWTH}, we display the growth rates for the kink mode in configurations given by Equation~(\ref{eq:pitch}) as a function of wavelength $\lambda_{\rm m=1}=2\pi/k_{\rm m=1}$ for varying pitch profiles. We indicate the critical length $L_0$ for which $q=1$ in Equation~(\ref{eq:safetyfactor}) by vertical lines. The actual system length then determines the instability growth. For wavelengths $L<L_0$, no unstable $m=1$ modes can be found. Therefore, the safety factor threshold of $q=1$ is a valid criterion for the onset of the kink instability. With flux tubes long enough to allow for unstable wavelengths to fit into the system, the fastest-growing mode with $L_0<\lambda_{\rm m=1}<L$ will dominate. The growth of the kink mode becomes faster for increasing pitch factor $\pitch_0$. These basic features hold for the different radial profiles of the inverse pitch with $\nu\in\left\{1,2\right\}$. However, the overall growth rates are significantly lower for larger $\nu$ (as a consequence of the nonlinear profile of the pitch parameter).

\subsubsection{The $m=2$ (fluting) mode}

We extend the instability analysis to the $m=2$ mode in Figure~\ref{fig:M2GROWTH}. Modes with $m>1$ grow with wavelengths shorter than the critical length $L_0$, below which the $m=1$ (kink) mode is suppressed. Again, their maximum growth rates increase for larger values of $\pitch_0$. As in the case of the $m=1$ (kink) mode, growth rates are significantly reduced for paraboloidal ($\nu = 2$) pitch profiles when compared to linear ($\nu = 1$) pitch profiles. Comparing the growth rates between the $m=1$ mode (Figure~\ref{fig:M1GROWTH}) and the $m=2$ mode (Figure~\ref{fig:M2GROWTH}), one finds ratios of $\text{Im}(\omega_{m=2})/\text{Im}(\omega_{m=1})=0.75-0.93$ for $\nu = 1$ and $\text{Im}(\omega_{m=2})/\text{Im}(\omega_{m=1})=0.58-0.72$ for $\nu = 2$. Although these measurements confirm the $m=1$ mode as the fastest-growing instability, the growth rate of the $m=2$ mode can become comparable. We evaluate the possibility of mode mixing in the following sections on simulated instability dynamics.

\section{Simulations}
\label{eq:sims}

We use FFE simulations to examine the instability growth in flux tubes described by Equation~(\ref{eq:pitch}). For this, we employ a high-order FFE method with optimized hyperbolic/parabolic cleaning parameters \citep{Mahlmann2020b,Mahlmann2020c,Mahlmann2021} that benefits from the \textsc{Carpet} driver \citep{Goodale2002a,Schnetter2004} and the \textsc{Einstein Toolkit} \citep{Loeffler2012,yosef_zlochower_2022_6588641}\footnote{\url{http://www.einsteintoolkit.org}}. FFE simulations integrate Maxwell's equations (\ref{eq:MAXI}-\ref{eq:MAXIV}) with currents set by the force-free condition~(\ref{eq:ffbalance}) and the constraints
\begin{align}
    \mathbf{E}\cdot\mathbf{B}=0&\qquad\text{(ideal fields)}\label{eq:FFI},\\
    E<B&\qquad\text{(magnetic dominance)}.\label{eq:FFII}
\end{align}
There are notable differences between FFE and the non-relativistic limit of MHD commonly used in laboratory and solar plasma physics. Only the drift velocity of frozen-in field lines is available in FFE; the plasma pressure and flow velocity are ordered out. Variations of the electric field can create local charge densities. Violations of the force-free constraints (\ref{eq:FFI}-\ref{eq:FFII}) rapidly dissipate nonideal electric fields. In FFE, the plasma modes reduce to Alfvénic and fast waves, both propagating with the speed of light and their characteristic polarizations \citep[see, e.g.,][]{Komissarov2002}. In this section, we track the displacement of magnetic field lines during various instabilities and quantify the induced dissipation.

The simulations fill a rectangular domain of size $x\times y\times z=\left[-4r_0,4r_0\right]\times \left[-4r_0,4r_0\right]\times \left[0,L\right]$ with resolution $\Delta x=\Delta y = \Delta z=r_0/N$. We choose $N=20$ as the number of grid points resolving the flux tube radius. The boundaries in the $xy$-directions are periodic. In the $z$-direction we use perfectly conducting surfaces with frozen-in field lines \citep[see][]{Munz2000,Mahlmann2023}. Simulations are initialized with background magnetic fields $B_z$ and $B_\phi$ as solutions to Equation~(\ref{eq:staticpitched}) determined by specifying the pitch profile in Equation~(\ref{eq:pitch}). The employed high-order FFE method suppresses discretization noises required to seed the instability growth. Therefore, we initialize the drift-velocity perturbation  
\begin{align}
    v^r = v_0\sin\left(k_zz\right)\sin\left(m\phi\right)f\left(r\right),
\end{align}
where $v_0/c =0.01$, $k_z=2\pi/L$, and $m=1$. In practice, this drift perturbation is set up by initializing electric fields according to Equation~(\ref{eq:idealeq}).

\subsection{Instability dynamics}
\label{sec:dynamics}

\begin{figure}
\centering
  \includegraphics[width=\linewidth]{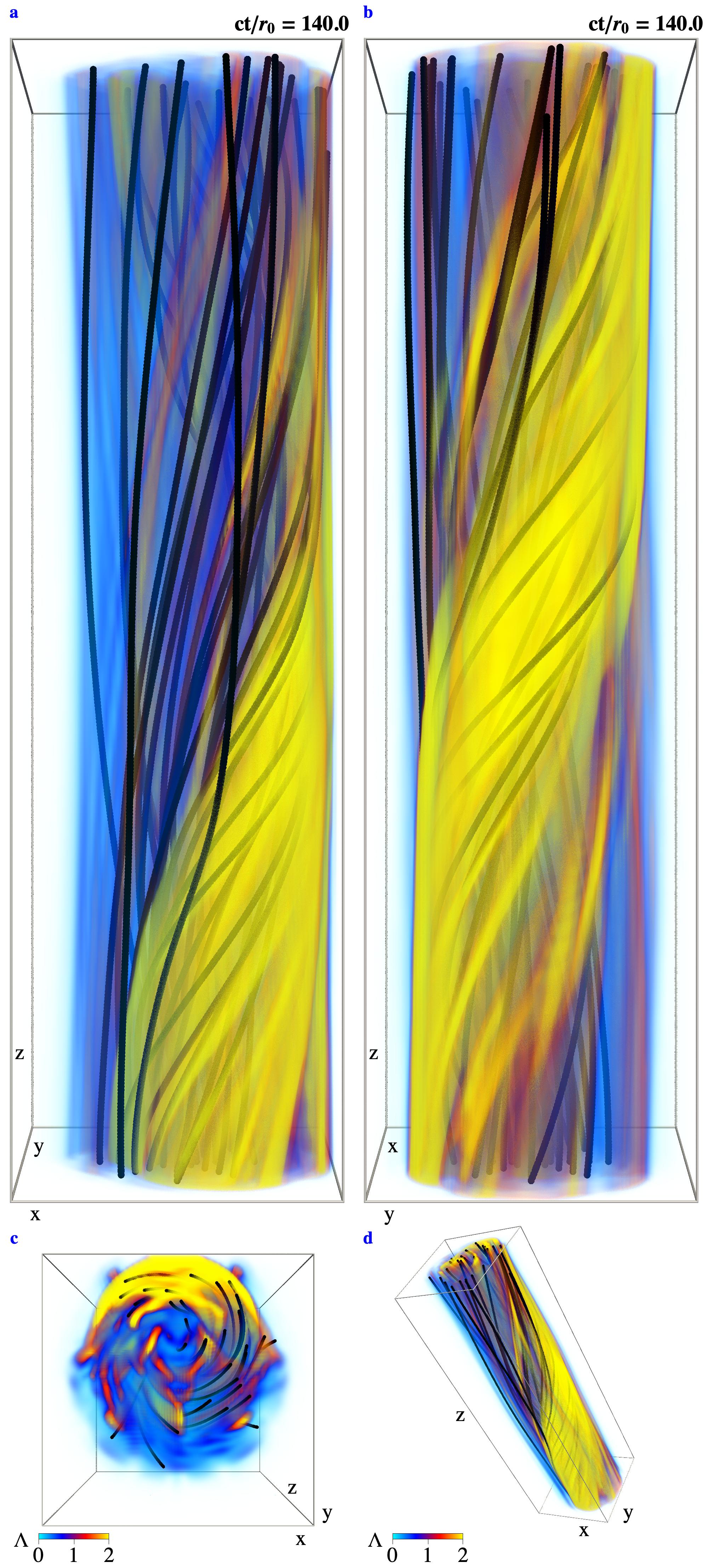}
  \caption{Three-dimensional visualization of the instability of a selected flux tube ($\nu = 1$, $\pitch_0 = 0.75$, $L/r_0 = 14$). We show views of the field-aligned current $\Lambda$ along the $x$-axis (a), the $y$-axis (b), the $z$-axis (c), and a diagonal view (d). In this configuration, the $m=1$ (kink) mode dominates, and strong currents wind around the initial flux tube center. Figure~\ref{fig:P075COMBI2} shows the time evolution of this setup.}
  \label{fig:L14_COMBI}
\end{figure}

\begin{figure}
\centering
  \includegraphics[width=\linewidth]{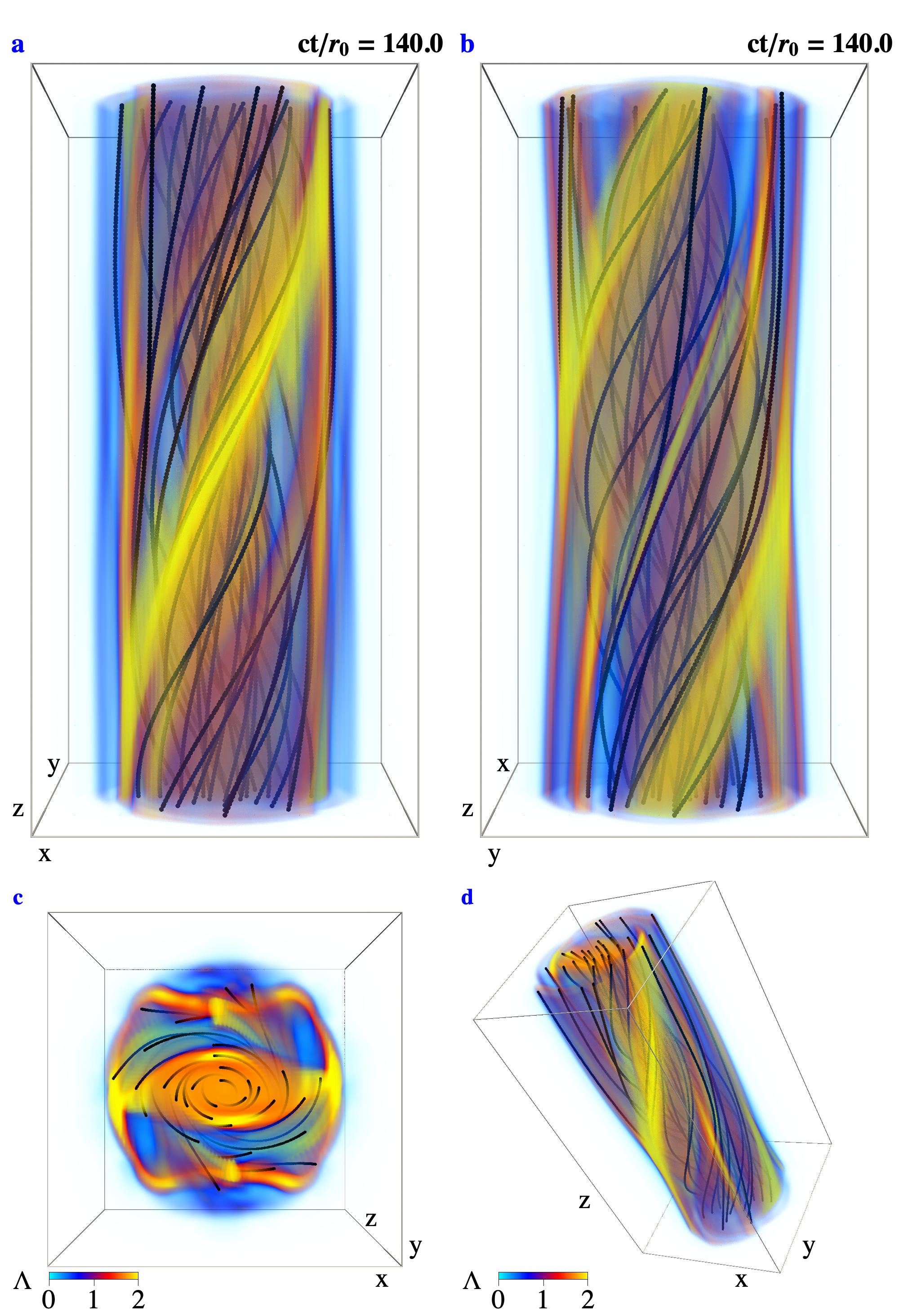}
  \caption{As Figure~\ref{fig:L14_COMBI} for a selected flux tube ($\nu = 1$, $\pitch_0 = 0.75$, $L/r_0 = 8$). In this configuration, the $m=1$ (kink) mode is suppressed, the $m=2$ (fluting) mode dominates, and strong currents quench around the initial flux tube center. Figure~\ref{fig:P075COMBI} shows the time evolution of this setup.}
  \label{fig:L8_COMBI}
\end{figure}

We first isolate the instability dynamics of force-free flux tubes for the $m=1$ (kink) and $m=2$ (fluting) modes. To this end, we follow the evolution of a setup with $\nu=1$ and $\pitch_0=0.75$ for two different flux tube lengths $a_0\equiv L/r_0=8$ and $L/r_0=14$. As shown in Figure~\ref{fig:M1GROWTH}, the kink mode of this configuration does not grow below $L/r_0\approx 8.4$, and the maximum growth rate of the $m=1$ mode is captured for $L/r_0\gtrsim 9.2$. Both configurations allow for the $m=2$ (fluting) mode to grow, as shown in Figure~\ref{fig:M2GROWTH}. We use the projection $\Lambda$ of the conserved force-free current along the magnetic field to visualize flux tubes in arbitrary field line geometries. This component of the current can be written in the form $\mathbf{j}_\parallel=\lambda\mathbf{B}$, with $\nabla\Lambda\cdot\mathbf{B}=0$; hence, $\Lambda$ is constant along magnetic field lines. 

Figure~\ref{fig:L14_COMBI} shows the field line configuration and currents after the onset of instability for $L/r_0=14$ (accompanied by significant dissipation of twist energy; see Section~\ref{sec:twistenergy}). The setup develops clear features of the $m=1$ (kink) mode, namely asymmetric variations of currents along the toroidal direction. The flux tube current cross section in panel c of Figure~\ref{fig:L14_COMBI} exhibits typical structures of the kink instability \citep[as discussed by][]{Davelaar2020,Mahlmann2023}.

The development of the kink mode is suppressed for the $L/r_0=8$ setup shown in Figure~\ref{fig:L8_COMBI}. The instability develops differently from the longer flux tube discussed above. Strong currents develop with an $m=2$ symmetry along the toroidal direction. The characteristic fluting manifests as a thinning of the flux tube in the $y$-direction with bulging along the $x$-direction. We note that both setups evaluated in this section ($L=8$ and $L=14$) become unstable at similar times. As established in Section~\ref{sec:instability}, the maximum growth rates of the $m=1$ and $m=2$ modes are comparable. We find $\text{Im}(\omega_{m=2})/\text{Im}(\omega_{m=1})\approx 0.048/0.052=0.92$ (see Figures~\ref{fig:M1GROWTH} and~\ref{fig:M2GROWTH}). We study the dependence of twist dissipation on the flux tube length and simultaneous growth of the kink and higher $m$ modes in the following section.

\subsection{Dissipation of twist energy}
\label{sec:twistenergy}

\begin{figure}
\centering
  \includegraphics[width=0.975\linewidth]{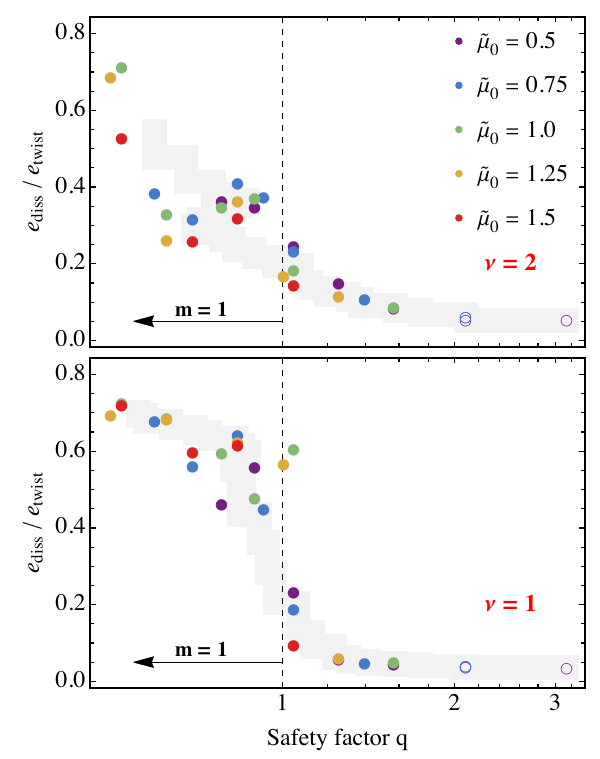}
  \caption{Energy dissipated during the evolution of perturbed flux tube equilibria for various pitch profiles with $\nu\in\left\{1,2\right\}$ and $\pitch_0\in\left\{0.5,0.75,1.0,1.25,1.5\right\}$. We display measurements of the dissipated energy $e_{\rm diss}/e_{\rm twist}$ and a moving average (gray line). Filled circles denote setups that develop instability; empty circles represent configurations that did not show any dissipation above the numerical diffusivity. The dashed vertical lines indicate the critical safety factor ($q=1$). The kink ($m=1$) mode will grow for setups located to the left of this line.}
  \label{fig:Dissipation}
\end{figure}

We scan the parameter space of pitch profiles by varying $\nu$ and $\pitch_0$ and measuring the dissipated magnetic energy $e_{\rm diss}$. We define the dissipated energy as the difference between twist energy before and after the development of instability, where the twist energy is
\begin{align}
e_{\rm twist}\left(t\right)=\int\text{d}V\,\frac{1}{2}\left[\mathbf{B}\left(t\right)-\mathbf{B}_{\rm bg}\right]^2.
\end{align}
We evolve perturbed equilibrium states in time for a duration of $ct/r_0 = 300-500$, making sure that the dynamical phase of the instability is fully captured and dissipation has returned to the low level of numerical diffusion. During the instability, twist energy is lost in steep gradients via numerical diffusion or by removing nonideal field components violating conditions (\ref{eq:FFI}-\ref{eq:FFII}). The total amount of dissipated energy is $e_{\rm diss}$. 

Figure~\ref{fig:Dissipation} displays the twist energy dissipation as a function of the safety factor $q$ for a set of $58$ different flux tubes. For large safety factors $q\gg 1$, no notable instability occurs, and dissipation is limited to numerical diffusion regardless of the pitch profile $\nu$ (empty circles). Sufficiently low safety factors $q\gtrsim 1$ allow for the growth of fluting ($m=2$) or higher-order modes. However, the dissipation in this region of the parameter space remains low with $e_{\rm diss}/e_{\rm twist}\lesssim 0.2$. For $q\lesssim 1$ the $m=1$ (kink) mode can grow, as was shown in Section~\ref{sec:instability}. The dissipation of twist energy jumps to larger values at $q\approx 1$, ranging between $e_{\rm diss}/e_{\rm twist}\approx 0.6$ for $\nu = 1$ and $e_{\rm diss}/e_{\rm twist}\approx 0.4$ for $\nu =2$.  For very low safety factors $q\ll 1$ the fraction of dissipated energy $e_{\rm diss}/e_{\rm twist}$ increases further to $e_{\rm diss}/e_{\rm twist}\approx 0.8$. 

Configurations with $q\lesssim 1$ develop both $m=1$ (kink) and $m=2$ (fluting) or higher-order modes. If the scale of the maximum growth rate of the kink instability is captured, the $m=1$ mode dominates in these cases. The growth of the instabilities quickly drives the system to an event with rapid dissipation and a rearrangement into a relaxed state of lower energy. Once the fluting instability develops for $q>1$, the system rearranges and dissipates energy equally fast. For $q\approx 1$, when the system size allows for the development of the $m=1$ mode but does not yet capture its maximum growth rate, the instability dynamics is more complex. First, $m=2$ or higher-order modes develop, driving the fluting of the flux tube and mild dissipation of twist energy. At later times, the $m=1$ (kink) instability significantly reduces the twist energy. Such events at the threshold of the critical safety factor with the subsequent development of modes of lower order can last three to five times longer than the cases of $q<1$ and $q>1$.

\section{Discussion}
\label{sec:discussion}

\begin{figure}
\centering
  \includegraphics[width=0.975\linewidth]{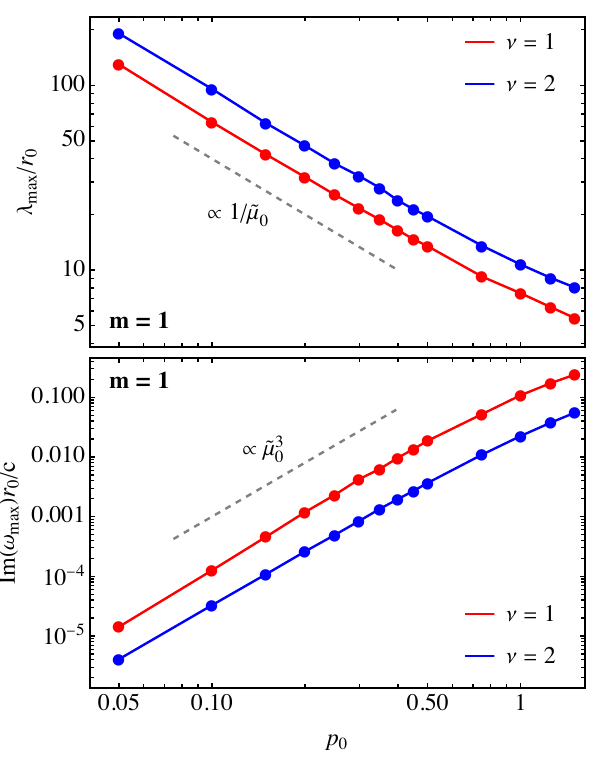}
  \caption{Maximum growth rate of the $m=1$ (kink) mode (bottom panel) and corresponding wavelength (top panel) for an extended range of pitch parameters $\pitch_0$. Configurations with a very small initial twist $\pitch_0\ll 1$ still show a growth of the kink mode. However, their maximum growth rate decays fast with $\text{Im}(\omega_{\rm max})r_0/c\propto \pitch_0^3$ (dashed gray line, bottom panel) for $\pitch_0\ll 1$, and the required system length for the $m=1$ (kink) mode scales with $\lambda_{\rm max}/r_0\propto 1/\pitch_0$ (dashed gray line, top panel).}
  \label{fig:MaxGrowthM1}
\end{figure}

\subsection{Scales and limits of the instability growth}
\label{sec:discussionscales}

We confirm in Section~\ref{sec:instability} that the $m=1$ (kink) mode is the fastest-growing instability of the system that also dissipates twist energy most efficiently (Section~\ref{sec:twistenergy}). Figure~\ref{fig:MaxGrowthM1} (bottom panel) extends the growth rate analysis shown in Section~\ref{sec:instability} for the $m=1$ (kink) mode to $\pitch_0=B_\phi/B_z\ll 1$. All probed configurations, even with a small initial twist,
have unstable solutions with a maximum growth rate
\begin{align}
   \frac{\text{Im}\left(\omega_{\rm max}\right)r_0}{c}\approx \eta \pitch_0^3\qquad\text{for}\, \pitch_0< 1.
   \label{eq:pitchscaling}
\end{align}
Here, $\eta$ is a parameter that depends on the radial pitch profile; it is obtained empirically as $\eta\approx 0.15$ for $\nu = 1$ and $\eta\approx 0.03$ for $\nu = 2$. We can estimate the time scale for the growth of instabilities as $t_i\equiv 1/\text{Im}(\omega_{\rm max})$:
\begin{align}
    t_i= 0.33\left(\frac{10^{-5}}{\text{Im}(\omega_{\rm max})r_0/c}\right)\left(\frac{r_0}{1\text{km}}\right)\text{s}.
    \label{eq:timescalegeneral}
\end{align}
By combining Equations~(\ref{eq:pitchscaling}) and~(\ref{eq:timescalegeneral}) we can estimate the growth rate of the kink instability close to the critical safety factor $q\approx 1$, where the flux tube aspect ratio is $a_0\equiv L_0/r_0\approx 2\pi/\pitch_0$:
\begin{align}
    t_i^{q\approx 1}\approx 4.0\times 10^{-6}a_0^3\left(\frac{0.15}{\eta}\right)^3\left(\frac{r_0}{1\text{km}}\right)\text{s}.
    \label{eq:growthtime}
\end{align}
The top panel of Figure~\ref{fig:MaxGrowthM1} confirms the relation $a_0\propto 1/\pitch_0$. The fastest-growing wavelength of the $m=1$ (kink) mode requires flux tubes with aspect ratios of up to $a_0>100$ for $\pitch_0\ll 1$. In other words, flux tubes have to be long and `skinny' to become unstable for small values of $\pitch_0$.

The instability analysis and simulations presented in this work have magnetic equilibria (Equation~\ref{eq:GSBasic}) as a starting point. However, the time scale of instability growth has to be compared to other characteristic time scales of the system. In realistic scenarios, one relevant scale is the rate at which a flux tube is twisted by footpoint motions in the line-tied boundaries. We denote the twisting time scale as $t_{\rm twist}$ and identify two relevant regimes: The slow-twisting regime with $t_i\ll t_{\rm twist}$, and the fast-twisting regime with $t_{\rm twist}\lesssim t_i$. In the slow twisting limit, the pitch parameter $\pitch_0$ of a flux bundle of fixed length $L$ will gradually increase until the system is disrupted by an $m=1$ or higher-order instability. As the configuration slowly approaches $q\gtrsim 1$, the dissipation during instability in this regime is likely low (see Section~\ref{sec:twistenergy} and Figure~\ref{fig:Dissipation}). In the fast-twisting regime, the inverse pitch $\pitch_0$ can increase beyond the critical value for a fixed system size $L$. The safety factor can, thus, reach $q\lesssim 1$ and significant dissipation of twist energy will likely occur (see Figure~\ref{fig:Dissipation}). Regardless of the safety factor, currents remain in the domain after instability and relaxation to a steady state. As we discuss in Appendix~\ref{app:relaxedstates}, the total currents of disrupted flux tubes drop less than their total twist magnetic fields. We interpret our findings in the context of different astrophysical environments in the following sections.

\begin{figure}
\centering
  \includegraphics[width=0.975\linewidth]{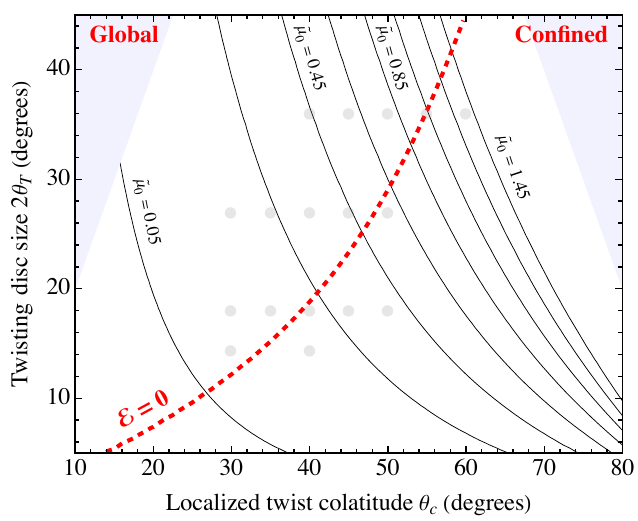}
  \caption{Estimate of a possible pitch parameter ($\pitch_0$) distribution for critical flux tubes ($q=1$) in a dipolar magnetar magnetosphere \citep[adapted from][with gray dots denoting the parameter space explored by their global magnetospheric simulations]{Mahlmann2023}. In a dipole field, flux tubes are parametrized by the center footpoint colatitude on the stellar surface $\theta_c$, and the angular extent of the twisting region $2\theta_T$. During the instability, the dipole magnetosphere can either open up in a large-scale eruption, or energy is dissipated locally \citep[transition roughly at the dashed red line, see][]{Mahlmann2023}. Critical flux tubes with $\pitch_0\ll 1$ require small $r_0$ and larger length $L$, and thus, footpoints closer to the poles.}
  \label{fig:Dipole}
\end{figure}

\subsection{Kink ($m=1$) instability in the magnetar corona}
\label{sec:discussionmagnetar}

Twisted magnetic fields likely play a key role in the region of active plasma processes in the relativistic magnetar magnetosphere, the so-called magnetar corona \citep[see, e.g.,][]{Beloborodov2007a,Beloborodov2013,Chen2017}. If their twist grows beyond a critical angle, the sheared dipole magnetosphere can erupt in flaring events with large-scale reconnection regions and energy dissipation \citep{Parfrey2013,Mahlmann2019,yuan2020,Sharma2023}. Three-dimensional flux tubes with twisting footpoints in a disk-like patch on the magnetar surface allow for rich eruption dynamics with lateral (torus-like) and helical (kink-like) instabilities \citep{Carrasco:2019aas,Mahlmann2023}. Figure~\ref{fig:Dipole} adapts a visualization of the threshold between large-scale global eruptions of the magnetosphere and confined eruptions by \citet[][red dashed line]{Mahlmann2023}. To connect the global context of the magnetar magnetosphere to the findings of this work, we display the magnetic pitch $\pitch_0$ for critical flux tubes ($q=1$) of different aspect ratios. The aspect ratio $a_0=L_0/r_0$ is inferred from the length of the center field line in a dipolar flux tube as well as the radius $r_0$ of the disk that induces foot point motion on the stellar surface \citep[see Figure 1 in][]{Mahlmann2023}. Flux tubes with low levels of inverse pitch require small diameters with large aspect ratios to become critical, $a_0=L_0/r_0=2\pi/\pitch_0$ for $q=1$. Shorter dipolar flux tubes located farther away from the poles need larger diameters and pitch parameters to become critical. For the simulation parameters scanned in \citet[][gray dots in Firgure~\ref{fig:Dipole}]{Mahlmann2023}, instabilities occur after $ct/L\approx 25$ for flux tubes closer to the poles ($\theta_c=30^\circ$) and $ct/L\approx 100$ for those closer to the equator ($\theta_c=60^\circ$). In combination with the angular dependence of the radial magnetic field in the dipole magnetosphere, $B_r= 2\mu_*\cos\theta/r^3$, this difference in time before eruption suggests a scaling in critical pitch that is consistent\footnote{Assuming $B_r(\theta_c=60^\circ)/B_r(\theta_c=30^\circ)\approx 0.57$ and $B_\phi(\theta_c=60^\circ)/B_\phi(\theta_c=30^\circ)\approx 4$, such that $\pitch(\theta_c=30^\circ)/\pitch(\theta_c=60^\circ)\approx 0.14$, roughly according to the difference in contours enclosing the gray dots in Figure~\ref{fig:Dipole}.} with the contours displayed in Figure~\ref{fig:Dipole}. 

Time scales of field line displacement on the magnetar surface are not yet well constrained. They range from slow quasi-steady shearing with $t_{\rm twist}$ on the order of years \citep[$\omega_s > 1\,\text{rad}\,\text{yr}^{-1}$, cf.][]{Parfrey2012} to rapid crust deformations with millisecond creep times during flares \citep[][]{Thompson2017,Thompson2022}. The slow shearing is clearly separated from the instability growth time $t_i$. However, for aspect ratios of $a_0 \gtrsim 6.3$, Equation~(\ref{eq:growthtime}) projects growth times above millisecond duration (also depending on the parameters $\eta$ and $r_0$). Thus, rapid crust deformations can reach the fast twisting regime with $t_{\rm twist}\lesssim t_i$. In this limit, the safety factor can reach $q<1$ due to the rapidly driven growth of pitch $\pitch_0$. As a consequence, significant part of the magnetospheric twist energy can dissipate (see Figure~\ref{fig:Dissipation}). We note that the simulations in \citet{Mahlmann2023} use $\omega_s < 1/25\times c/R_*$, equating to $t_{\rm twist}>5.2\times 10^{-3}\left(R_*/10\text{km}\right)\text{s}$. For large aspect ratios, or low pitch parameters $\pitch_0$, this choice of twist time scale allows for $t_{\rm twist}\lesssim t_i$ and possibly enhances magnetospheric dissipation. We acknowledge the limitation of such direct comparison between the straight flux tubes discussed in this work and the dipolar magnetosphere in Section~\ref{sec:limitations}.

\subsection{Mixed instabilities in magnetized coronae}
\label{sec:discussionmixed}

In the case of flux bundles twisting in the slow limit of $t_i\ll t_{\rm twist}$ or if the twist injection ceases while $q>1$, $m=2$ (fluting) and higher-order modes can develop. As we demonstrate in Section~\ref{sec:instability}, the $m=2$ mode grows slower than the $m=1$ mode, though their growth rates are in general comparable. In the dynamic phase of the instability, this coincidence of growth rates manifests by mixing of the symmetric, short-wavelength fluting and the asymmetric, long-wavelength kinking patterns (see Appendix~\ref{app:relaxedstates}). We find systems that are only susceptible to the $m=2$ mode to dissipate a comparably small fraction of the twist energy (Figure~\ref{fig:Dissipation}) and maintain significant currents after relaxation (Appendix~\ref{app:relaxedstates}). However, the fluting develops with its short wavelengths and can potentially drive turbulence even when the kink mode is present. While the $m=1$ (kink) mode develops predominantly around the center of the flux tube, the $m=2$ (fluting) mode drives the dynamics at resonant surfaces in the outer layers. 

The possibility of mode mixing in flux tubes was discussed, though not observed, in the context of the solar corona by \citet{Quinn2022}. Instead of perturbing a equilibrium configuration with a specific profile that seeds the growth of an $m=1$ (kink) mode, \citet{Quinn2022} rely on noise introduced by the continuous twisting of a flux tube to drive the instability. With this strategy, which is somewhat closer to the realistic flux tube evolution, they allow for the development of $m>1$ modes that are usually not addressed in the literature.\footnote{We note that the numerical work conducted for this paper initially followed a similar approach. We especially considered the late-stage evolution of flux tubes where the continuous motion of footpoints was turned off after an initial twisting episode. In such setups, we found kink-like and fluting-like dynamics, seeded purely by the boundary and developing different levels of dissipation. The long evolution times until instability onset made the systematic study of such setups prohibitive for a large parameter space.} \citet{Quinn2022} consider the role of nonideal effects for the flux tube dynamics in MHD and find that the cumulative ohmic heating is mainly driven by the kink instability. The analysis we present in Figure~\ref{fig:Dissipation} equally suggests that most dissipation occurs during the development of the $m=1$ (kink) mode. 

In contrast to \citet{Quinn2022} we do not find a significantly delayed onset of the kink mode due to the growth of $m=2$ (fluting) patterns. This may be due to the absence of resistive/viscous effects in FFE, which effectively propagates all perturbations at the fastest velocity, the speed of light. Still, for configurations erupting at $q\approx 1$ we find a notable interaction between the $m=1$ and $m=2$ modes (see Section~\ref{sec:instability}). With the growth of the kink mode suppressed initially, such systems exclusively develop $m=2$ (fluting) dynamics. Once the configuration is no longer in equilibrium, the flux tube changes its properties, and the $m=1$ (kink) mode can grow regardless of its suppression in the initial state. Thus, for line-tied relativistic flux tubes of critical length with $q\gtrsim 1$ the $m=2$ mode can drive the system to rapid dissipation by the kink instability. Some of the configurations studied in this work develop higher-order, vortex-like structures in the nonlinear phase of the instability (see, e.g., Figures~\ref{fig:P075COMBI2} and~\ref{fig:P125COMBI}). Such `fingers' also appear in other studies of the relativistic kink instability \citep[e.g.,][]{Bromberg2019,Davelaar2020}; their dispersion and role for dissipation in the nonlinear phase should be evaluated further.


\subsection{Limitations}
\label{sec:limitations}

The presented simulations of flux tube dynamics use the force-free limit of ideal MHD. Such FFE models are not suitable to capture the conversion of magnetic energy consistently due to the absence of information about plasma inertial properties like particle number densities or nonideal electric fields \citep[see][]{Mahlmann2020c,Mahlmann2021}. The measurements of dissipation shown in Section~\ref{sec:twistenergy}, especially its time scales, should therefore be understood as the energy loss in the limit of rapid cooling of nonideal fields. How a magnetosphere with line-tied shear fills with plasma and how active plasma processes can dissipate the injected twist even without the onset of large-scale instabilities was previously studied in reduced dimensionality \citep[e.g.,][]{Beloborodov2007a}. Its consistent plasma dynamics for realistic coronal geometries have to be further evaluated by particle-in-cell methods.

After a disruption by an instability, especially those occurring close to the critical value of $q\approx 1$, some twist energy and currents remain in the system (see Figure~\ref{fig:Dissipation} and Appendix~\ref{app:relaxedstates}). A continuing twist of the flux tube footpoints could drive further eruptions of configurations with larger and larger twist energies \citep[as modeled by][]{Mahlmann2023}. However, in such secondary and later events, the flux tubes are no longer in an axisymmetric equilibrium as considered in this work. Calculating the safety factor and corresponding instability criteria in nonaxisymmetric states is less straightforward. It will require careful consideration of the flux tube geometry, as in \citet{Stefanou2022} who derive generalized force-free Grad-Shafranov equilibria of magnetospheres with nontrivial twisted flux ropes. We defer studying the dynamic (in)stability of such configurations to future work.

The instability analysis presented in Section~\ref{sec:instability} can be improved, particularly regarding the neglect of line-tying on the perturbation $\xi$. Contrary to typical flux-tube analyses in the solar corona, we do not impose any $z$-dependent boundary conditions on $\xi$. Line-tying with $\xi$ vanishing smoothly at field line footpoints \citep[e.g.,][for the photosphere]{Hood1979,Hood1981} could change the derived instability threshold and its dependence on the system size \citep[cf.][]{Goedbloed1994}. Specifically, an eigenvalue analysis including $z$-dependent boundaries may yield different outcomes than the purely radial one-dimensional balance equation discussed in Section~\ref{sec:instability}. The effects of $z$-dependent boundaries on the stability of specific flux-tube geometries will be explored in future work. The full 3D simulations discussed in Section~\ref{eq:sims} use perfect conductor boundaries, an instantaneous line-tying. We find that the dynamics in these simulations are in good agreement with the predictions of unstable wavelengths derived by the instability analysis (Section~\ref{sec:instability}). For small line-tying scale heights, the system length is well defined, and the one-dimensional analysis in Section~\ref{sec:instability} provides good insights, for example, estimates of the minimum system length or the maximum instability growth rate for a given pitch factor. To produce more complete dispersion relations, future instability analyses in FFE could adopt innovative frameworks like \texttt{Legolas} \citep{Claes2020}. Such solvers systematically analyze the eigensystem of linearized MHD and can be extended to capture nonideal effects \citep{DeJonghe2022,dejonghe2023}.

Finally, this work focuses on cylindrical flux tubes and disregards any curvature effects experienced by bent structures commonly observed in the solar corona and expected, for instance, around magnetars and magnetized accretion flows. We only use the presented findings to \emph{qualitatively} supplement models that take into account coronal geometries. MHD modes were analyzed in geometries relevant to laboratory plasmas early on, most notably in extensions to toroidal geometries \citep{Goedbloed1975}. In parallel to the vast progress in tokamak applications, MHD models with coronal geometries built up a track record in the solar physics community \citep[e.g.,][]{Amari2003,Gerrard2004,Torok2005,Torok2010,Gordovskyy2011,Gordovskyy2014,Pinto2016,Ripperda2017,Ripperda2017a,Sauppe2018}. There are only a few comparable works for global magnetospheric instabilities around compact objects \citep[e.g.,][]{Carrasco:2019aas,Mahlmann2023,Most2023} that have to be studied in greater detail in the future.

\section{Conclusion}
\label{sec:conclusions}

This paper examines the behavior of line-tied flux tubes with an axial twist of $\pitch = B_\phi / B_z$ in the force-free regime. We examine perturbations of force-free flux tube equilibria both analytically (Section~\ref{sec:instability}) and in FFE simulations (Section~\ref{sec:dynamics}). Depending on the flux tube parameters, our analysis predicts and shows the development of the kink ($m=1$) and/or fluting ($m=2$) instabilities. To test the flux tubes' susceptibility to the $m=1$ mode, we apply a stability indicator called the safety factor $q = 2 \pi r_0 \pitch/ L$, which represents the inverse of the number of azimuthal windings of the magnetic field along the flux tube length. Resulting analyses of growth timescale and energy dissipation are then applied to astrophysical systems such as magnetars and the lower solar corona.

The safety factor is the key criterion to determine if line-tied FFE flux tubes without initial field line motion become kink unstable (see Figure~\ref{fig:M1GROWTH}). We find rapid growth of the kink instability for $q<1$. For $q>1$, fluting ($m=2$) and higher $m$ modes may develop when the flux tube is long enough to accommodate these modes. For $q\gtrsim 1$, an initial deformation of the flux tube by fluting modes can catalyze the development of the kink instability (Section~\ref{sec:dynamics}). 

We tested two flux tube pitch profiles $\pitch(r) \propto \pitch_0 r^\nu$ in this work, where $\nu \in \{1,2\}$. Theoretically, we find the $\nu =1$ case generates larger growth rates for both the kink and fluting instabilities. The maximum growth rate of the fluting mode is within 60–90\% of the kink mode (Figures~\ref{fig:M1GROWTH} and~\ref{fig:M2GROWTH}), resulting in instabilities that evolve on similar timescales. However, the fluting mode dissipates only about 20\% of the initial twist energy, while the kink mode dissipates between 40\% and 80\% (Figure~\ref{fig:Dissipation}). The maximum growth rate of the kink mode, $\text{Im}(\omega_{\rm max})r_0 / c$,  scales as $\eta \pitch_0^3$, where $\eta$ is found experimentally to be 0.03–0.15 depending on the pitch profile (Figure~\ref{fig:MaxGrowthM1}). The wavelength corresponding to this fastest-growing $m=1$ (kink) mode, $\lambda_{\rm max} / r_0$, is proportional to $\pitch_0^{-1}$. 

The explosive release of magnetic energy in magnetospheres, including around magnetized compact objects like magnetars and black hole accretion systems, can be driven by the instability of twisted magnetic flux bundles. Similar to CMEs of the Sun, twisted flux tubes anchored to a neutron star or accretion disk can erupt and dissipate magnetic energy when their twist exceeds a critical value. In nature, the twist in a flux tube is likely established by footpoint motions of magnetic field lines in the line-tied boundary. In this paper, we develop an intuition for the onset of instabilities in line-tied force-free flux tubes and the dissipation associated with such events. We suggest that fast footpoint shearing at the line-tied boundary can tap the regime of efficient dissipation during $m=1$ (kink) instabilities ($q<1$). However, if the shear builds up slowly compared to the growth of the kink mode, it is likely that higher-order instabilities distort the flux tube with only little magnetospheric dissipation. This by itself can drive flux tubes to states with localized $q\lesssim 1$ regions though the safety factor is not straightforwardly obtained in nonaxisymmetric configurations. Our simulations and growth rate analysis confirm that flux tubes of any pitch value $B_z/B_\phi$ can become kink unstable given sufficient length such that $q\lesssim 1$ (see Section~\ref{sec:discussion}). The idealized analysis and simulation of isolated line-tied magnetic flux bundles presented in this paper provide an intuition for the dynamics of flux tubes and the corresponding dissipation limits that can be used in the complex modeling of radiative plasma processes around magnetars and magnetized accretion disks.

\section*{Acknowledgements}
We thank R. Keppens for peer-reviewing this work and improving it through thoughtful and educational comments. The authors thank B. Crinquand, A. Philippov, and E. Sobacchi for useful discussions. We are grateful for the funding provided through NASA grant 80NSSC18K1099. This research was facilitated by the Multimessenger Plasma Physics Center (MPPC), NSF grant PHY-2206607. NR acknowledges the Department of Energy (DOE) support for the Summer Undergraduate Laboratory Internship (SULI) program at the Princeton Plasma Physics Laboratory (PPPL). Work at PPPL is supported under US DOE Contract No. DE-AC02-09CH11466. JFM acknowledges support by the National Science Foundation under grants No. AST-1909458 and AST-1814708. The presented numerical simulations were enabled by
the \emph{MareNostrum} supercomputer (Red Española de Supercomputación, AECT-2022-3-0010), and the \emph{Stellar}
cluster (Princeton Research Computing).

\bibliography{literature.bib}

\newpage
\appendix

\section{Instability evolution and relaxed states}
\label{app:relaxedstates}

\begin{figure}
\centering
  \includegraphics[width=\linewidth]{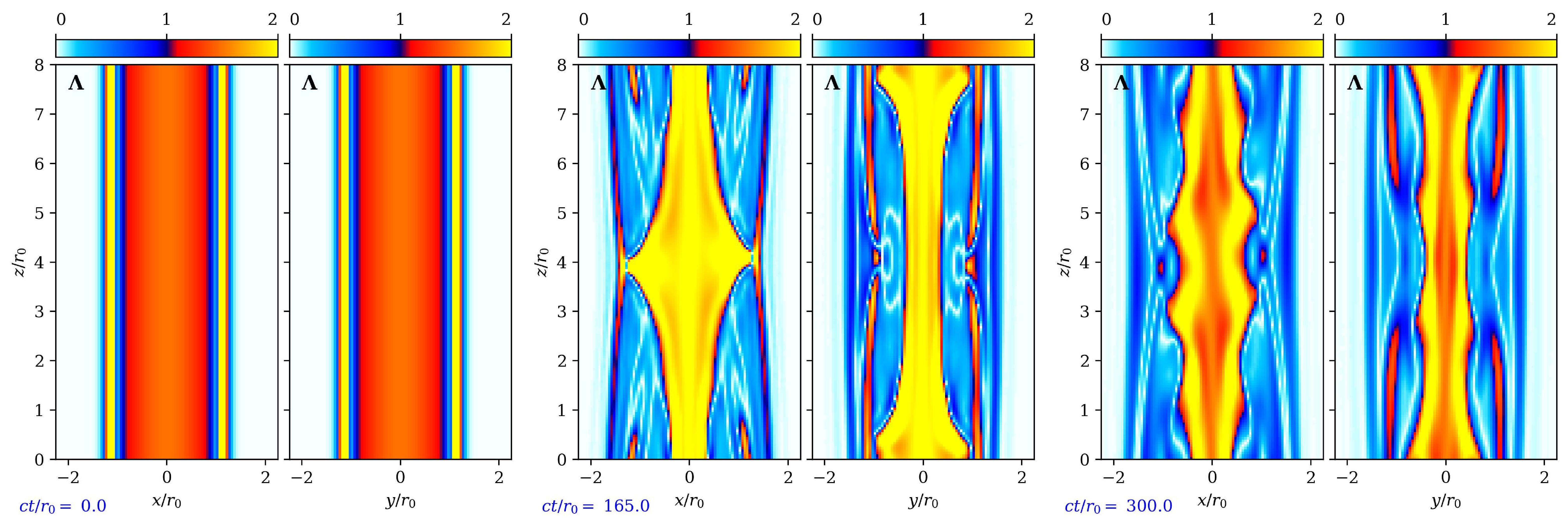}
  \caption{Currents during the evolution of a selected flux tube ($\nu = 1$, $\pitch_0 = 0.75$, $L/r_0 = 8$). The middle panel shows structures that are characteristic of the $m=2$ (fluting) mode. Significant currents remain in the relaxed state (right panel). We provide an animated version of this figure as supplementary material \citep{SupplementaryMediaC}. It shows the evolution during times $ct/r_0=0$ to $300$; the real time duration of the animation is $5\,{\rm s}$. Reactions of the flux tube to the instability are perceivable at $ct/r_0=75$ (real time: $1\,{\rm s}$), the most dynamic phase is at $ct/r_0=150$ (real time: $3\,{\rm s}$).}
  \label{fig:P075COMBI}
\end{figure}

\begin{figure}
\centering
  \includegraphics[width=\linewidth]{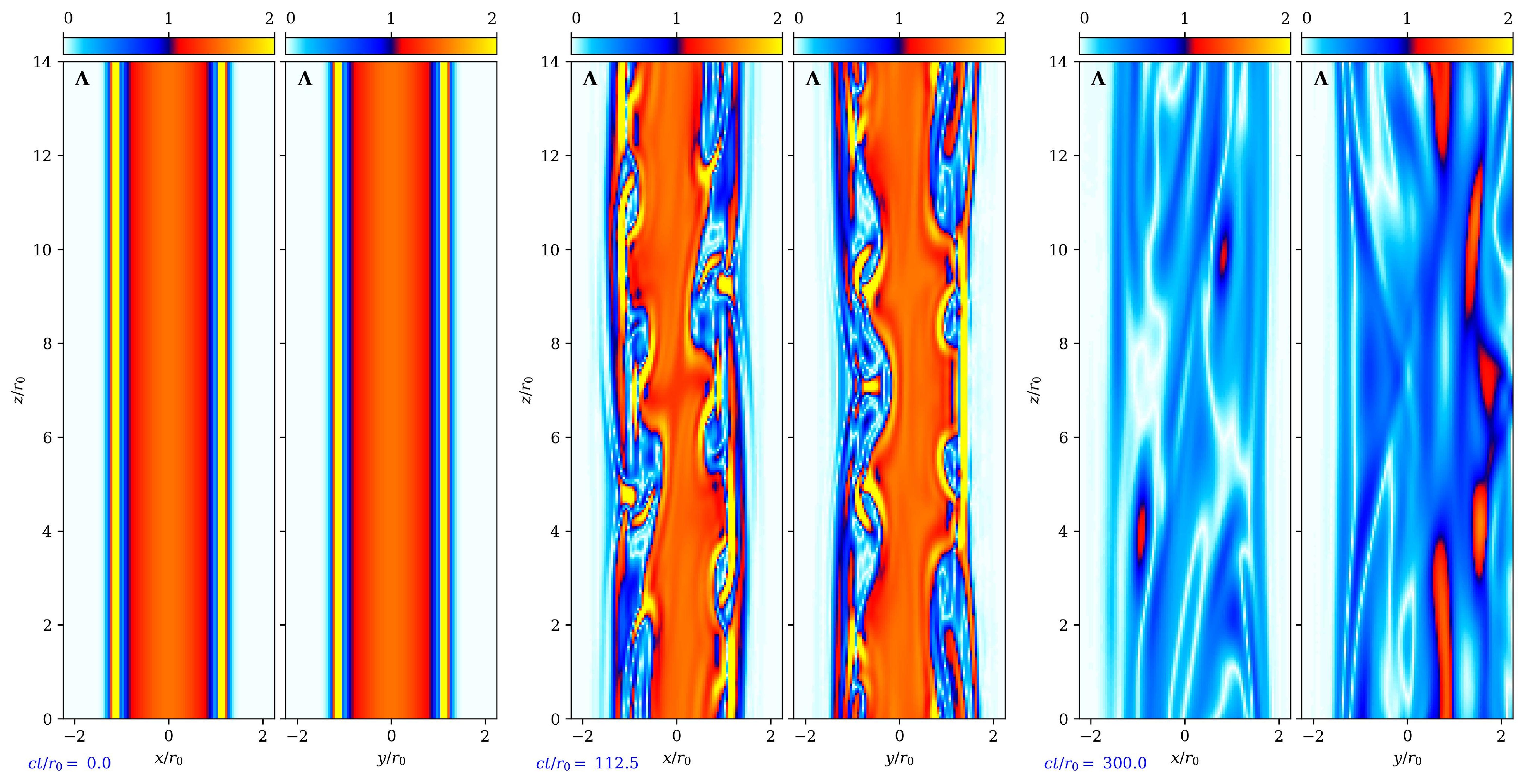}
  \caption{Currents during the evolution of a selected flux tube ($\nu = 1$, $\pitch_0 = 0.75$, $L/r_0 = 14$). The middle panel shows structures that are characteristic of the asymmetric $m=1$ (kink) mode as well as the $m=2$ (fluting) mode. The $m=2$ mode has shorter wavelengths than the $m=1$ mode, as discussed in Section~\ref{sec:instability}. In the relaxed state (right panel) some currents remain though currents are weaker than in the dynamic phase of shorter configurations (cf. Figure~\ref{fig:P075COMBI}). We provide an animated version of this figure as supplementary material \citep{SupplementaryMediaA}. It shows the evolution during times $ct/r_0=0$ to $300$; the real time duration of the animation is $5\,{\rm s}$. Reactions of the flux tube to the instability are perceivable at $ct/r_0=80$ (real time: $1\,{\rm s}$), the most dynamic phase is at $ct/r_0=115$ (real time: $2\,{\rm s}$).}
  \label{fig:P075COMBI2}
\end{figure}

\begin{figure}
\centering
  \includegraphics[width=\linewidth]{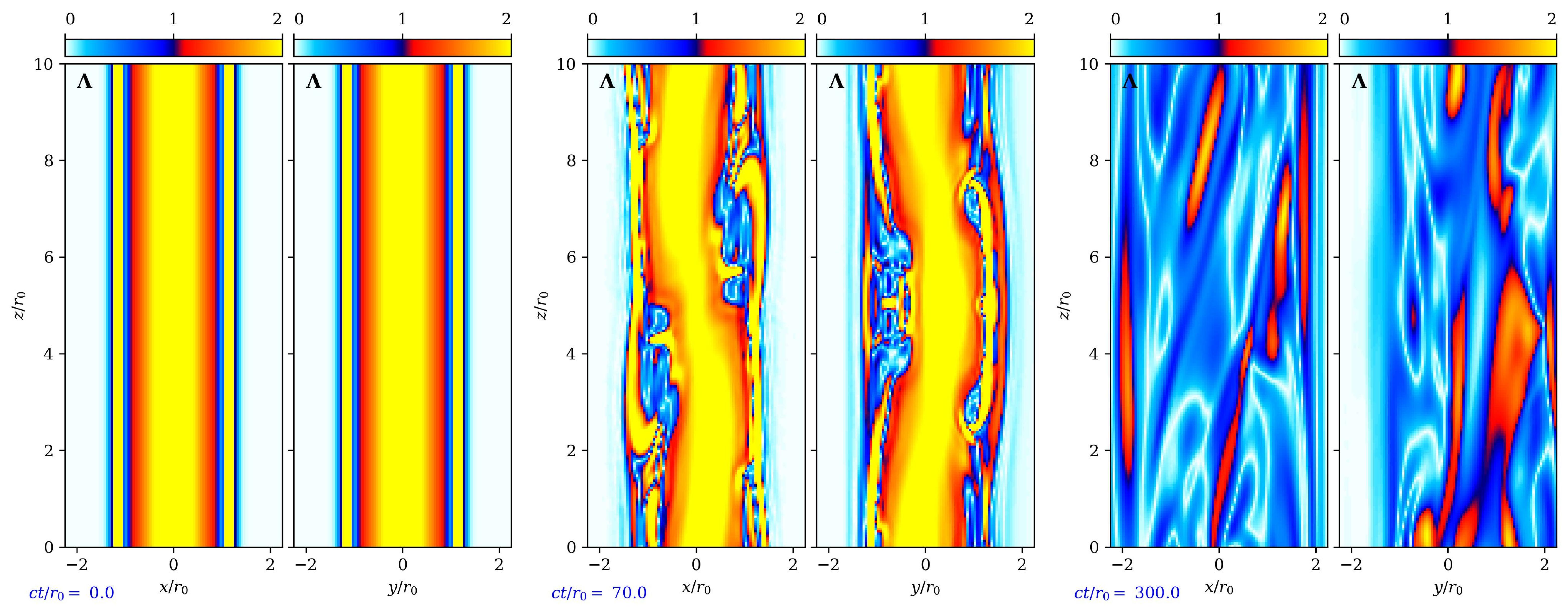}
  \caption{Currents during the evolution of a selected flux tube ($\nu = 1$, $\pitch_0 = 1.25$, $L/r_0 = 10$). The middle panel shows structures that are characteristic of the dominating $m=1$ (kink). Some currents remain in the relaxed state (right panel). We provide an animated version of this figure as supplementary material \citep{SupplementaryMediaB}. It shows the evolution during times $ct/r_0=0$ to $300$; the real time duration of the animation is $5\,{\rm s}$. Reactions of the flux tube to the instability are perceivable at $ct/r_0=60$ (real time: $1\,{\rm s}$), the most dynamic phase is at $ct/r_0=70$ (real time: $1\,{\rm s}$).}
  \label{fig:P125COMBI}
\end{figure}

It is instructive to follow the dynamics of the conserved force-free current $\lambda$ throughout the instability development. As discussed in Section~\ref{sec:instability}, the length and pitch profile of flux tubes determines the dominantly growing modes and sets the amount of dissipated twist energy $e_{\rm diss}/e_{\rm twist}$ (Section~\ref{sec:twistenergy}). Short flux tubes with a safety factor $q>1$ suppress the kink mode (see Figure~\ref{fig:M1GROWTH}). higher-order modes, like the $m=2$ (fluting) mode, can still grow in such systems. The middle panel of Figure~\ref{fig:P075COMBI} shows a clear characteristic of the symmetric $m=2$ (fluting) mode for a selected configuration with $\nu =1$, $\pitch_0=0.75$, and $L=8$. Specifically, the flux tube develops a bulging in the $xz$-plane and thinning in the $yz$-plane. Increasing the length of the same pitch configuration such that $q\lesssim 1$ allows the growth of the $m=1$ (kink) mode, as shown in Figure~\ref{fig:P075COMBI2}. The middle panel of Figure~\ref{fig:P075COMBI2} shows the symmetric, short-wavelength bulging and thinning of the $m=2$ (fluting) mode, as well as the asymmetric, long-wavelength runaway pattern of the $m=1$ (kink) mode. This characteristic shape of the $m=1$ (kink) mode becomes more prominent for configurations with larger initial pitch parameter $\pitch_0$, like the $\nu =1$, $\pitch_0=1.25$, $L=10$ setup shown in Figure~\ref{fig:P125COMBI}.

\begin{figure}
\centering
  \includegraphics[width=0.85\linewidth]{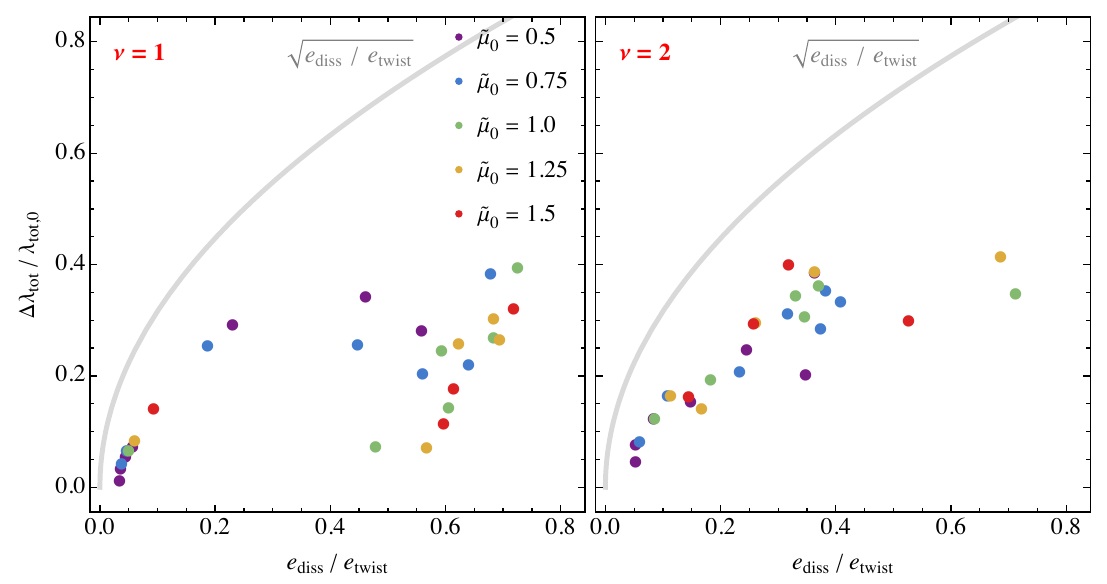}
  \caption{Change in total current $\Delta\lambda_{\rm tot}/\lambda_{\rm tot,0}$ against dissipated energy $e_{\rm diss}/e_{\rm twist}$ for the models described in Section~\ref{sec:twistenergy}.}
  \label{fig:DISSVSLAMBDA}
\end{figure}

With the dissipation of twist energy during the development of instabilities (see Figure~\ref{fig:Dissipation}), the considered systems relax to a new steady state. In general, this relaxed state is no longer axiymmetric or solved by the force balance in Equation~(\ref{eq:GSBasic}). As shown in the right panels of Figures~\ref{fig:P075COMBI} to~\ref{fig:P125COMBI}, currents remain in the final states of the evolution. Configurations with suppressed $m=1$ (kink) modes maintain significant and ordered currents in the relaxed state (Figure~\ref{fig:P075COMBI}). However, systems subject to the $m=1$ (kink) instability end up with lower levels of current that extend further than the initial flux tube boundary with patches of stronger currents (Figures~\ref{fig:P075COMBI2} and~\ref{fig:P125COMBI}). We quantify the loss of current during the instability by measuring the change in total current $\lambda_{\rm tot}=\int\lambda\,\text{d}V$ in the domain. Figure~\ref{fig:DISSVSLAMBDA} shows the relative difference of initial and final currents $\Delta\lambda_{\rm tot}/\lambda_{\rm tot,0}$ as a function of dissipated twist energy $e_{\rm diss}/e_{\rm twist}$ for all models in Section~\ref{sec:twistenergy}. These measurements demonstrate that currents persist for all configurations in the final state of evolution. The change in total current is below the change of total twist magnetic fields, with $\Delta\lambda_{\rm tot}/\lambda_{\rm tot,0}<\sqrt{e_{\rm diss}/e_{\rm twist}}$. 

\end{document}